\documentclass{ws-book9x6}
\usepackage{ws-book-har}
\title{Information and Computation}

\makeindex
\begin{document}

\chapter{On the Algorithmic Nature\\ \text{ }of the World}
\author[Hector Zenil and Jean-Paul Delahaye]{Hector Zenil$^\ddagger$\footnote{hector.zenil@lifl.fr} and Jean-Paul Delahaye$^\dagger$\footnote{delahaye@lifl.fr}}
%\index[aindx]{Author, F.}  % or \aindx{Author, F.}
%\index[aindx]{Author, S.} % or \aindx{Author, S.}

%\usepackage{setspace} 
%\usepackage[dvipdfm]{graphicx}
%\usepackage{graphics}
%\usepackage{graphicx}
%\usepackage{amsmath}
%\usepackage{amsfonts}
%\usepackage{amssymb}

\address{$^\dagger$$^\ddagger$Laboratoire d'Informatique Fondamentale de Lille (USTL/CNRS)\\$^\ddagger$Institut d'Histoire et de Philosophie des Sciences et des Techniques\\(CNRS/Paris I/ENS)}

\section{Abstract}

We propose a test based on the theory of algorithmic complexity and an experimental evaluation of Levin's universal distribution to identify evidence in support of or in contravention of the claim that the world is algorithmic in nature. To this end we have undertaken a statistical comparison of the frequency distributions of data from physical sources on the one hand--repositories of information such as images, data stored in a hard drive, computer programs and DNA sequences--and the frequency distributions generated by purely algorithmic means on the other--by running abstract computing devices such as Turing machines, cellular automata and Post Tag systems. Statistical correlations were found and their significance measured.

\section{Introduction}

A statistical comparison has been undertaken of the frequency distributions of data stored in physical repositories on the one hand--DNA sequences, images, files in a hard drive--and of frequency distributions produced by purely algorithmic means on the other--by running abstract computational devices like Turing machines, cellular automata and Post Tag systems.

A standard statistical measure is adopted for this purpose. The Spearman rank correlation coefficient quantifies the strength
of the relationship between two variables, without making any prior assumption as to the particular nature of the relationship between them.

\subsection{Levin's Universal Distribution}

Consider an unknown operation generating a binary string of length $k$ bits. If the method is uniformly random, the probability of finding a particular string $s$ is exactly $2^{-k}$, the same as for any other string of length $k$. However, data is usually produced not at random but by a process. There is a measure which describes the expected output frequency distribution of an abstract machine running a program. A process that produces a string $s$ with a program $p$ when executed on a universal Turing machine $T$ has probability $m(s)$. As $p$ is itself a binary string, $m(s)$ can be defined as being the probability that the output of a universal prefix Turing machine $T$ is $s$ when provided with a sequence of fair coin flip inputs interpreted as a program. Formally,

\begin{equation*}
m(s) = \Sigma_{T(p) = s} 2^{-|p|}
\end{equation*}

where the sum is over all halting programs $p$ for which $T$ outputs the string $s $,  with $|p|$ the length of the program $p$. As $T$ is a prefix universal Turing machine, the set of valid programs forms a prefix-free set\footnote{No element is a prefix of any other, a property necessary to keep $0<m(s)<1$ for all $s$ and therefore a valid probability measure} and thus the sum is bounded due to Kraft's inequality. For technical details see \cite{calude, li, downey}.

Formulated by Leonid Levin\cite{levin}, $m$ has many remarkable properties\cite{kirchherr}. It is closely related to the concept of algorithmic complexity\cite{chaitin1} in that the largest value of the sum of programs is dominated by the shortest one, so one can actually write $m(s)$ as follows:

\begin{equation*}
m(s) = 2^{-K(s)+O(1)}
\end{equation*}

In a world of computable processes, $m(s)$ establishes that simple patterns which result from simple processes are likely, while complicated patterns produced by complicated processes (long programs) are relatively unlikely.

It is worth noting that, unlike other probability measures, $m$ is not only a probability distribution establishing that there are some objects that have a certain probability of occurring according to said distribution, it is also a distribution specifying the order of the particular elements  in terms of their individual algorithmic complexity.

\section{The null hypothesis}

When looking at a large-enough set of data following a distribution, one can in statistical terms safely assume  that the source generating
the data is of the nature that the distribution suggests. Such is the case when a set of data follows, for example, a Gaussian distribution, where depending on certain statistical variables, one can say with a high degree of certitude that the process generating the data is of a random nature.

When observing the world, the outcome of a physical phenomenon $f$ can be seen as the result of a natural process $P$. One may ask how the probability distribution of a set of process of the type of $P$ looks like.

If one would like to know whether the world is algorithmic in nature one would need first to tell how an algorithmic world would look like. To accomplish this, we've conceived and performed a series of experiments to produce data by purely algorithmic means in order to compare sets of data produced by several physical sources. At the right level a simplification of the data sets into binary language seems always possible. Each observation can measure one or more parameters (weight, location, etc.) of an enumeration of independent distinguishable values, a discrete sequence of values\footnote{This might be seen as an oversimplification of the concept of a natural process and of its outcome when seen as a binary sequence, but the performance of a physical experiment always yields data written as a sequence of individual observations as a valid sample of certain phenomena.}.

If there is no bias in the sampling method or the generating process itself and no information about the process is known, the principle of indifference\cite{thompson}\footnote{also known as principle of insufficient reason.} states that if there are $n>1$ possibilities mutually exclusive, collectively exhaustive and only distinguishable for their names then each possibility should be assigned a probability equal to $1/n$ as the simplest non-informative prior. The null hypothesis to test is that the frequency distributions studied herein from several different independent sources are closer to the experimental calculation of Levin's universal distribution than to the uniform (simplest non-informative prior) distribution. To this end average output frequency distributions by running abstract computing devices such as cellular automata, Post tag systems and Turing machines were produced on the one hand, and by collecting data to build distributions of the same type from the physical world on the other.

\subsection{Frequency distributions}

The distribution of a variable is a description of the relative number of times each possible outcome occurs in a number of trials. One of the most common probability distributions describing physical events is the normal distribution, also known as the Gaussian or Bell curve distribution, with values more likely to occur due to small random variations around a mean.

There is also a particular scientific interest in power-law distributions, partly from the ease with which certain general classes of mechanisms generate them. The demonstration of a power-law relation in some data can point to specific kinds of mechanisms that might underlie the natural phenomenon in question, and can indicate a connection with other, seemingly unrelated systems.

As explained however, when no information is available, the simplest distribution is the uniform distribution, in which values are equally likely to occur. In a macroscopic system at least, it must be assumed that the physical laws which govern the system are not known well enough to predict the outcome. If one does not have any reason to choose a specific distribution and no prior information is available, the uniform distribution is the one making no assumptions according to the principle of indifference. This is supposed to be the distribution of a balanced coin, an unbiased die or a casino roulette where the probability of an outcome $k_i$  is $1/n$ if $k_i$ can take one of $n$ possible different outcomes.

\subsection{Computing abstract machines}

An abstract machine consists of a definition in terms of input, output, and the set of allowable operations used to turn the input into the output. They are of course algorithmic by nature (or by definition). Three of the most popular models of computation in the field of theoretical computer science were resorted to   produce data of a purely algorithmic nature: these were deterministic Turing machines (denoted by $TM$), one-dimensional cellular automata (denoted by $CA$) and Post Tag systems ($TS$).

The Turing machine model represents the basic framework underlying many concepts in computer science, including the definition of algorithmic complexity cited above. The cellular automaton is a well-known model which, together with the Post Tag system model, has been studied since  the foundation of the field of abstract computation by some of its first pioneers. All three models are Turing-complete. The descriptions of the models follow formalisms used in \cite{wolfram}. 

\subsubsection{Deterministic Turing machines}

The Turing machine description consists of a list of rules (a finite program) capable of manipulating a linear list of cells, called the \emph{tape}, using an access pointer called the \emph{head}. The finite program can be in any one of a finite set of states $Q$ numbered from $1$ to $n$, with $1$ the state at which the machine starts its computation. Each tape cell can contain $0$ or $1$ (there is no special blank symbol). Time is discrete and the steps are ordered from $0$ to $t$ with $0$ the time at which the machine starts its computation. At any given time, the head is positioned over a particular cell and the finite program starts in the state $1$. At time $0$ all cells contain the same symbol, either $0$ or $1$. A rule $i$ can be written in a $5$-tuple notation as follows $\{s_i,k_i,s_i^\prime,k_i^\prime,d_i\}$, where $s_i$ is the tape symbol the machine's head is scanning at time $t$,  $k_i$ the machine's current 'state' (instruction) at time $t$, $s_i^\prime$ a unique symbol to write (the machine can overwrite a 1 on a 0, a 0 on a $1$, a $1$ on a $1$, or a $0$ on a $0$) at time $t+1$, $k_i^\prime$ a state to transition into (which may be the same as the one it was already in) at time $t+1$, and $d_i$ a direction to move in time $t+1$, either to the right ($R$) cell or to the left ($L$) cell, after writing. Based on a set of rules of this type, usually called a transition table, a Turing machine can perform the following operations: 1. write an element from $A=\{0,1\}$, 2. shift the head one cell to the left or right, 3. change the state of the finite program out of $Q$. When the machine is running it executes the above operations at the rate of one operation per step. At a time $t$ the Turing machine produces an output described by the contiguous cells in the tape visited by the head.

Let $T(0),T(1),\ldots,T(n),\ldots$ be a natural recursive enumeration of all $2$-symbol deterministic Turing machines. One can, for instance, begin enumerating by number of states, starting with all $2$-state Turing machines, then $3$-state, and so on. Let $n$, $t$ and $k$ be three integers. Let $s(T(n), t)$ be the part of the contiguous tape cells that the head visited after $t$ steps. Let's consider all the $k$-tuples, i.e. all the substrings of length $k$ from $s(T(n), t)=\{s_1,s_2, \ldots, s_u\}$, i.e. the following $u-k+1$ $k$-tuples: $\{(s_1, \ldots , s_k) ,(s_2, \ldots , s_{k+1}), \ldots, (s_{u-k+1}, \ldots , s_u)\}$.

Now let $N$ be a fixed integer. Let's consider the set of all the $k$-tuples produced by the first $N$ Turing machines according to a recursive enumeration after running for $t$ steps each. Let's take the count of each $k$-tuple produced.

From the count of all the $k$-tuples, listing all distinct strings together with their frequencies of occurrence, one gets a probability distribution over the finite set of strings in $\{0,1\}^k$.

For the Turing machines the experiments were carried out with $2$-symbol $3$-state Turing machines. There are $(4n)^{2n}$ possible different $n$-state $2$ symbol Turing machines according to the $5$-tuple rule description cited above. Therefore $(4 \times 3)^{(2 \times 3)}=2985984$ $2$-symbol $3$-state Turing machines. A sample of $2000$ $2$-symbol $3$-state Turing machines was taken. Each Turing machine's runtime was set to $t=100$ steps starting with a tape filled with $0s$ and then once again with a tape filled with $1s$ in order to avoid any undesired asymmetry due to a particular initial set up.

\subsubsection{One-dimensional Cellular Automata}

An analogous standard description of one-dimensional $2$-color cellular automata was followed. A one-dimensional cellular automaton is a collection of cells on a row that evolves through discrete time according to a set of rules based on the states of neighboring cells that are applied in parallel to each row over time. When the cellular automaton starts its computation, it applies the rules at a first step $t=0$. If $m$ is an integer, a neighborhood of $m$ refers to the cells on both sides, together with the central cell, that the rule takes into consideration at row $t$ to determine the value of a cell at the step $t+1$. If $m$ is a fraction of the form $p/q$, then $p-1$ are the cells to the left and $q-1$ the cells to the right taken into consideration by the rules of the cellular automaton.

For cellular automata, the experiments were carried out with $3/2$-range neighbor cellular automata starting from a single $1$ on a background of $0s$ and then again starting from a single $0$ on a background of $1s$ to avoid any undesired asymmetry from the initial set up. There are $2^{2m+1}$ possible states for the cells neighboring a given cell ($m$ at each side plus the central cell), and two possible outcomes for the new cell; there are therefore a total of $2^{2^{2m+1}}$ one-dimensional $m$-neighbor $2$-color cellular automata, hence $2^{2^{(2 \times 3/2)+1}}=65536$ cellular automata with rules taking two neighbors to the left and one to the right. A sample of $2000$ $3/2$-range neighbor cellular automata was taken.

As for Turing machines, let $A(1),A(2),\ldots, A(n),\ldots$ be a natural recursive enumeration of one dimensional $2$-color cellular automata. For example, one can start enumerating them by neighborhood starting from range $1$ (nearest-neighbor) to $3/2$-neighbor, to $2$-neighbor and so on, but this is not mandatory.

Let $n$, $t$ and $k$ be three integers. For each cellular automaton $A(n)$, let $s(A(n),t)$ denote the output of the cellular automata defined as the contiguous cells of the last row produced after $t=100$ steps starting from a single black or white cell as described above, up to the length that the scope of the application of the rules starting from the initial configuration may have reached (usually the last row of the characteristic cone produced by a cellular automaton). As was done for Turing machines, tuples of length $k$ were extracted from the output.

\subsubsection{Post Tag Systems}

A Tag system is a triplet $(m, A, P)$, where $m$ is a positive integer, called the deletion number. $A$ is the alphabet of symbols (in this paper a binary alphabet). Finite (possibly empty) strings can be made of the alphabet $A$. A computation by a Tag system is a finite sequence of strings produced by iterating a transformation, starting with an initially given initial string at time $t=0$. At each iteration $m$ elements are removed from the beginning of the sequence and a set of elements determined by the production rule $P$ is appended onto the end, based on the elements that were removed from the beginning. Since there is no generalized standard enumeration of Tag systems\footnote{To the authors' knowledge.}, a random set of rules was generated, each rule having equal probability. Rules are bound by the number of $r$ elements (digits) on the left and right hand blocks of the rule.  There are a total of ${(k^r+1-1)/(k-1)}^{k^n}$ possible rules if blocks up to length $r$ can be added at each step. For $r = 3$, there are $50625$ different 2-symbol Tag systems with deletion number 2. In this experiment, a sample of $2000$ 2-Tag systems (Tag systems with deletion number $2$) were used to generate the frequency distributions of Tag systems to compare with. 

An example of a rule is $\{0 \rightarrow 10, 1 \rightarrow 011, 00 \rightarrow \epsilon, 01 \rightarrow 10, 10 \rightarrow 11\}$, where no term on any side has more than $3$ digits and there is no fixed number of elements other than that imposed to avoid multiple assignations of a string to several different, i.e. ambiguous, rules. The empty string  $\epsilon$ can only occur among the right hand terms of the rules. The random generation of a set of rules yields results equivalent to those obtained by following and exhausting a natural recursive enumeration.

As an illustration\footnote{For illustration only no actual enumeration was followed.}, assume that the output of the first $4$ Turing machines following an enumeration yields the output strings $01010$, $11111$, $11111$ and $01$ after running $t=100$ steps. If $k=3$, the tuples of length $3$ from the output of these Turing machines are: $((010,101,010), (111,111,111),(111,111,111))$; or grouped and sorted from higher to lower frequency: $(111,010,101)$ with frequency values 6, 2, and 1 respectively. The frequency distribution is therefore $((111, 2/3), (010, 2/9),(101, 1/9))$, i.e. the string followed by the count divided by the total. If the strings have the same frequency value they are lexicographically sorted.

\begin{figure}
\begin{center}
\includegraphics[width=1\textwidth]{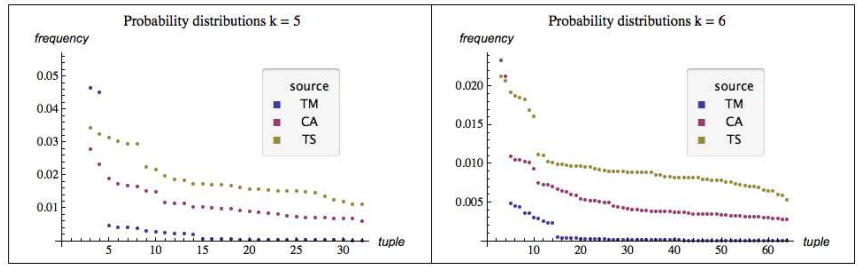}
\end{center}
\caption{The output frequency distributions from running abstract computing machines. The $x$-axis shows all the $2^k$ tuples of length $k$ sorted from most to least frequent. The $y$-axis shows the frequency values (probability between $0$ and $1$) of each tuple on the $x$-axis.}
\label{graphA}
\end{figure}

The output frequency distributions produced by abstract machines as described above are evidently algorithmic by nature (or by definition), and they will be used both to be compared one to each other, and to the distributions extracted from the physical real world.

\subsection{Physical sources}

Samples from physical sources such as DNA sequences, random images from the web and data stored in a hard drive were taken and transformed into data of the same type (i.e. binary tuples) of the produced by the abstract computing machines. We proceeded as follows: a representative set of random images was taken from the web using the random image function available online at the Wikipedia Creative Commons website at \emph{http://commons.wikimedia.org/wiki/Special:Random/File} (as of May, 2009), none of them larger than 1500 linear pixels\footnote{The sum of the width and height.} to avoid any bias due to very large images. All images were transformed using the $Mathematica$ function $Binarize$ that converts multichannel and color images into black-and-white images by replacing all values above a globally determined threshold. Then all the $k$-tuples for each row of the image were taken and counted, just as if they had been produced by the abstract machines from the previous section.

Another source of physical information for comparative purposes  was  a random selection of human gene sequences of Deoxyribonucleic acid (or simply DNA). The DNA was extracted from a random sample of 100 different genes (the actual selection is posted in the project website cited in section \ref{conclusions}).

There are four possible encodings for translating a DNA sequence into a binary string using a single bit for each letter: \{$G \rightarrow 1, \text{ } T \rightarrow 1, \text{ } C \rightarrow 0, \text{ } A \rightarrow 0$\}, \{$G \rightarrow 0, \text{ } T \rightarrow 1, \text{ } C \rightarrow 0, \text{ } A \rightarrow 1$\}, \{$G \rightarrow 1, \text{ } T \rightarrow 0, \text{ } C \rightarrow 1, \text{ } A \rightarrow 0$\}, \{$G \rightarrow 0, \text{ } T \rightarrow 0, \text{ } C \rightarrow 1, \text{ } A \rightarrow 1$\}.

To avoid any artificially induced asymmetries due to the choice of a particular encoding, all four encodings were applied to the same sample to build
a joint frequency distribution. All the $k$-tuples were counted and ranked likewise.

There might be still some biases by sampling genes rather than sampling DNA segments because genes might be seen as conceived by researchers to focus on functional segments of the DNA. We've done however the same experiments taking only a sample of a sample of genes, which is not a gene by itself but a legitimate sample of the DNA, producing the same results (i.e. the distributions remain stable). Yet, the finding of a higher \emph{algorithmicity} when taking gene samples as opposed to DNA general sampling might suggest that effectively there is an embedded encoding for genes in the DNA as functional subprograms in it, and not a mere research convenience.

A third source of information from the real world was a sample of data contained in a hard drive. A list of all the files contained in the hard drive was generated using a script, and a sample of $100$ files was taken for comparison, with none of the files being greater than 1 Mb in order to avoid any bias due to a very large file. The stream was likewise cut into k-tuples, counted and ranked to produce the frequency distribution, as  for DNA. The count of each of the sources yielded a frequency distribution of $k$-tuples (the binary strings of length $k$) to compare with.

One may think that data stored in a hard drive already has a strong algorithmic component by the way that it has been produced (or stored in a digital computer) and therefore it makes no or less sense to compare with to any of the algorithmic or empirical distributions. It is true that the data stored in a hard drive is in the middle of what we may consider the abstract and the physical worlds, which makes it however interesting as an experiment by its own from our point of view. But more important, data stored in a hard drive is of very different nature, from text files subject to the rules of language, to executable programs, to music and video, all together in a single repository. Hence, it is not obvious at all why a frequency distribution from such a rich source of different kind of data might end up resembling to other distributions produced by other physical sources or by abstract machines.

\subsection{Hypothesis testing}

The frequency distributions generated by the different sources were statistically compared to look for any possible correlation. A correlation test was carried out and its significance measured to validate either the null hypothesis or the alternative (the latter being that the similarities are due to chance).

Each frequency distribution is the result of the count of the number of occurrences of the k-tuples from which the binary strings of length $k$ were extracted. Comparisons were made with $k$ set from $4$ to $7$.

\subsubsection{Spearman's rank correlation coefficient}

The Spearman rank correlation coefficient\cite{snedecor} is a non-parametric measure of correlation that makes no assumptions about
the frequency distribution of the variables. Spearman's rank correlation coefficient is equivalent to the Pearson correlation on ranks. Spearman's rank correlation coefficient is usually denoted by the Greek letter $\rho$.

The Spearman rank correlation coefficient is calculated as follows:

\begin{equation*}
\rho = 1 - ((6 \sum d_i^2)/(n(n^2 - 1)))
\end{equation*}

\noindent where $d_i$ is the difference between each rank of corresponding values of $x$ and $y$, and $n$ the number of pairs of values.

Spearman's rank correlation coefficient can take real values from -$1$ to $1$, where -$1$ is a perfect negative (inverse) correlation, $0$ is no correlation and $1$ is a perfect positive correlation.

The approach to testing whether an observed $\rho$ value is significantly different from zero, considering the number of elements, is to calculate the probability that it would be greater than or equal to the observed $\rho$ given the null hypothesis using a permutation test\cite{good} to ascertain that the obtained value of $\rho$ obtained is unlikely to occur by chance (the alternative hypothesis). The common convention is that if the value of $\rho$ is between $0.01$ and $0.001$ the correlation is strong enough, indicating that the probability of having found the correlation is very unlikely to be a matter of  chance, since it would occur one time out of hundred (if closer to $0.01$) or a thousand (if closer to $0.001$), while if it is between $0.10$ and $0.01$ the correlation is said to be weak, although yet quite unlikely to occur by chance, since it would occur one time out of ten (if closer to $0.10$) or a hundred (if closer to $0.01$)\footnote{Useful tables with the calculation of levels of significance for different numbers of ranked elements are available online (e.g. at \emph{http://www.york.ac.uk/depts/\\ maths/histstat/tables/spearman.ps} as of May 2009).}. The lower the significance level, the stronger the evidence in favor of the null hypothesis. Tables \ref{table1}, \ref{table2}, \ref{table3} and \ref{table4} show the Spearman coefficients between all the distributions for a given tuple length $k$. 

\begin{table}[htdp]
\begin{center}
\textsc{\small{Cellular automata distribution} \text{  } \text{  } \text{  } \small{Hard \text{  } drive \text{  } distribution}  \text{  } \text{  } \\}
\begin{tabular}{|c|c|c|}
\hline
$rank$ & \it{string} (s) & $count$ \\
& & $(pr(s))$ \\
\hline
1 & 1111 & .35 \\
2 & 0000 & .34 \\
3 & 1010 & .033 \\
4 & 0101 & .032 \\
5 & 0100 & .026 \\
6 & 0010 & .026 \\
7 & 0110 & .025 \\
8 & 1011 & .024 \\
9 & 1101 & .023 \\
10 & 1001 & .023 \\
11 & 0011 & .017 \\
12 & 0001 & .017 \\
13 & 1000 & .017 \\
14 & 1100 & .016 \\
15 & 0111 & .016 \\
16 & 1110 & .017 \\
\hline
\end{tabular}
\text{  } \text{  } \text{  }
\begin{tabular}{|c|c|c|}
\hline
$rank$ & \it{string} (s) & $count$  \\
& & $(pr(s))$ \\
\hline
1 & 1111 & .093 \\
2 & 0000 & .093 \\
3 & 1110 & .062 \\
4 & 1000 & .062 \\
5 & 0111 & .062 \\
6 & 0001 & .062 \\
7 & 0100 & .06 \\
8 & 0010 & .06 \\
9 & 1101 & .06 \\
10 & 1011 & .06 \\
11 & 1100 & .056 \\
12 & 0011 & .056 \\
13 & 1001 & .054 \\
14 & 0110 & .054 \\
15 & 1010 & .054 \\
16 & 0101 & .054 \\
\hline
\end{tabular}
\end{center}
\caption{Examples of frequency distributions of tuples of length $k=4$, one from random files contained in a hard drive and another produced by running cellular automata. There are $2^4=16$ tuples each followed by its count (represented as a probability value between 0 and 1).}
\label{comparison}
\end{table}

\begin{table}[t]
\tbl{Spearman coefficients for $K=4$. Coefficients indicating a significant correlation are indicated by $\dagger$ while correlations with higher significance are indicated with $\ddagger$.}
{\begin{tabular}{@{}|c|c|c|c|c|c|c|@{}}
\hline
 \text{k = 4} & \text{HD} & \text{ADN} & \text{IMG} & \text{TM} & \text{CA} & \text{TS} \\
\hline
 \text{HD} & 1$\ddagger$ & 0.67$\ddagger$ & 0.4 & 0.29 & 0.5 & 0.27 \\
 \text{DNA} & 0.67$\ddagger$ & 1$\ddagger$ & 0.026 & 0.07 & 0.39$\dagger$ & 0.52$\dagger$ \\
 \text{IMG} & 0.4$\dagger$ & 0.026 & 1$\ddagger$ & 0.31 & 0.044 & 0.24 \\
 \text{TM} & 0.29 & 0.07 & 0.31 & 1$\ddagger$ & 0.37$\dagger$ & 0.044 \\
 \text{CA} & 0.5$\dagger$ & 0.39$\dagger$ & 0.044 & 0.37 & 1$\ddagger$ & 0.023 \\
 \text{TS} & 0.27 & 0.52$\dagger$ & 0.24 & 0.044 & 0.023 & 1$\ddagger$ \\
\hline
\end{tabular}
}
\begin{tabnote}
%$^{\text a}$Sample table footnote.\\
%$^{\text b}$Another sample table footnote.
\end{tabnote}
\label{table1}
\end{table}

\begin{table}[t]
\tbl{Spearman coefficients for $K=5$.}
{\begin{tabular}{@{}|c|c|c|c|c|c|c|@{}}
\hline
 \text{k = 5} & \text{HD} & \text{ADN} & \text{IMG} & \text{TM} & \text{CA} & \text{TS} \\
\hline
 \text{HD} & 1$\ddagger$ & 0.62$\ddagger$ & 0.09 & 0.31$\dagger$ & 0.4$\ddagger$ & 0.25$\dagger$ \\
 \text{ADN} & 0.62$\ddagger$ & 1$\ddagger$ & 0.30 & 0.11 & 0.39$\dagger$ & 0.24$\dagger$ \\
 \text{IMG} & 0.09 & 0.30$\dagger$ & 1$\ddagger$ & 0.32$\dagger$ & 0.60$\ddagger$ & 0.10 \\
 \text{TM} & 0.31$\dagger$ & 0.11 & 0.32$\dagger$ & 1$\ddagger$ & 0.24$\dagger$ & 0.07 \\
 \text{CA} & 0.4$\ddagger$ & 0.39$\dagger$ & 0.24$\dagger$ & 0.30$\dagger$ & 1$\ddagger$ & 0.18 \\
 \text{TS} & 0.25$\dagger$ & 0.24$\dagger$ & 0.10 & 0.18 & 0.021 & 1$\ddagger$ \\
 \hline
\end{tabular}
}
\begin{tabnote}
%$^{\text a}$Sample table footnote.\\
%$^{\text b}$Another sample table footnote.
\end{tabnote}
\label{table2}
\end{table}

\begin{table}[t]
\tbl{Spearman coefficients for $K=6$.}
{\begin{tabular}{@{}|c|c|c|c|c|c|c|@{}}
\hline
 \text{k = 6} & \text{HD} & \text{ADN} & \text{IMG} & \text{TM} & \text{CA} & \text{TS} \\
\hline
 \text{HD} & 1$\ddagger$ & 0.58$\ddagger$ & 0& 0.27$\dagger$ & 0.07 & 0.033 \\
 \text{DNA} & 0.58$\ddagger$ & 1$\ddagger$ & 0& 0.12 & 0.14 & 0\\
 \text{IMG} & 0& 0& 1$\ddagger$ & 0.041 & 0.023 & 0.17$\dagger$ \\
 \text{TM} & 0.27$\dagger$ & 0.12 & 0.041 & 1$\ddagger$ & 0& 0\\
 \text{CA} & 0.07 & 0.14 & 0.023 & 0& 1$\ddagger$ & 0.23$\dagger$ \\
 \text{TS} & 0.033 & 0& 0.17$\dagger$ & 0& 0.23$\dagger$ & 1$\ddagger$ \\
 \hline
\end{tabular}
}
\begin{tabnote}
%$^{\text a}$Sample table footnote.\\
%$^{\text b}$Another sample table footnote.
\end{tabnote}
\label{table3}
\end{table}

\begin{table}[t]
\tbl{Spearman coefficients for $K=7$.}
{\begin{tabular}{@{}|c|c|c|c|c|c|c|@{}}
\hline
 \text{k = 7} & \text{HD} & \text{ADN} & \text{IMG} & \text{TM} & \text{CA} & \text{TS} \\
\hline
 \text{HD} & 1$\ddagger$ & 0& 0.091$\dagger$ & 0.073 & 0& 0.11$\dagger$ \\
 \text{DNA} & 0& 1$\ddagger$ & 0.07 & 0.028 & 0.12 & 0.019 \\
 \text{IMG} & 0.091$\dagger$ & 0.07 & 1$\ddagger$ & 0.08$\dagger$ & 0.15$\ddagger$ & 0\\
 \text{TM} & 0.073 & 0.028 & 0.08$\dagger$ & 1$\ddagger$ & 0.03 & 0.039 \\
 \text{CA} & 0& 0.12 & 0.15$\ddagger$ & 0.03 & 1$\ddagger$ & 0\\
 \text{TS} & 0.11$\dagger$ & 0.019 & 0& 0.039 & 0& 1$\ddagger$ \\
 \hline
\end{tabular}
}
\begin{tabnote}
%$^{\text a}$Sample table footnote.\\
%$^{\text b}$Another sample table footnote.
\end{tabnote}
\label{table4}
\end{table}

When graphically compared, the actual frequency values of each tuple among the $2^k$ unveil a correlation in values along different distributions. The $x$ and $y$ axes are in the same configuration as in the graph \ref{graphA}: The $x$-axis plots the $2^k$ tuples of length $k$ but unlike the graph \ref{graphA} they are lexicographically sorted (as the result of converting the binary string into a decimal number). The table \ref{tablestrings} shows this lexicographical order as an illustration for $k=4$. The $y$-axis plots the frequency value (probability between $0$ and $1$) for each tuple on the $x$-axis.

\begin{table}[t]
\tbl{Illustration of the simple lexicographical order of the $2^4$ tuples of length $k = 4$ as plotted in the $x$-axis.}
{\begin{tabular}{@{}|c|c|c|c|@{}}
 \hline
$x$-axis order & tuple & $x$-axis order & tuple \\
 \hline
 0 & 0000 & 8 & 1000 \\
 1 & 0001 &  9 & 1001 \\
 2 & 0010 & 10 & 1010 \\
 3 & 0011 &  11 & 1011 \\
 4 & 0100 &  12 & 1100 \\
 5 & 0101 & 13 & 1101 \\
 6 & 0110 & 14 & 1110 \\
 7 & 0111 &  15 & 1111 \\
\hline
\end{tabular}
}
\begin{tabnote}
%$^{\text a}$Sample table footnote.\\
%$^{\text b}$Another sample table footnote.
\end{tabnote}
\label{tablestrings}
\end{table}

\begin{figure}
\begin{center}
\includegraphics[width=1\textwidth]{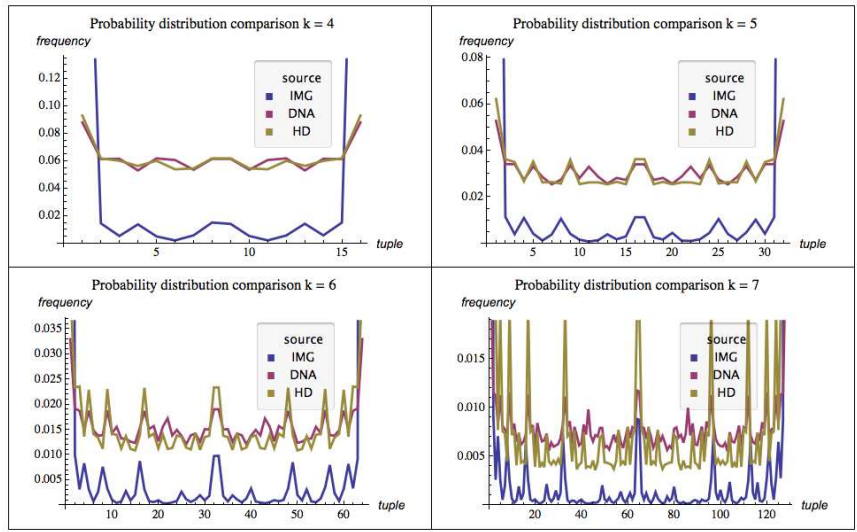}
\end{center}
\caption{Frequency distributions of the tuples of length $k$ from physical sources: binarized random files contained in a hard drive (HD), binarized sequences of Deoxyribonucleic acid (DNA) and binarized random images from the world wide web. The data points have been joined for clarity.}
\label{graph1}
\end{figure}

\begin{figure}
\begin{center}
\includegraphics[width=1\textwidth]{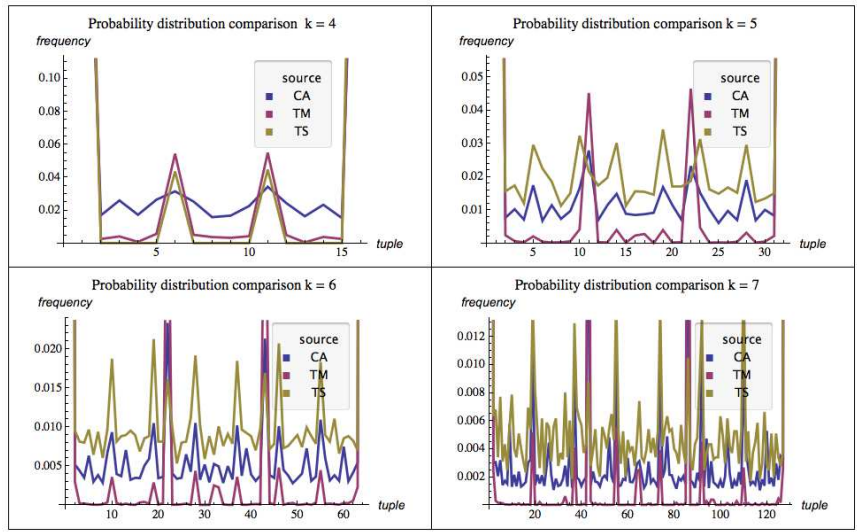}
\end{center}
\caption{Frequency distributions of the tuples of length $k$ from abstract computing machines: deterministic Turing machines (TM), one-dimensional cellular automata (CA) and Post Tag systems (TS).}
\label{graph2}
\end{figure}

\begin{figure}
\begin{center}
\includegraphics[width=1\textwidth]{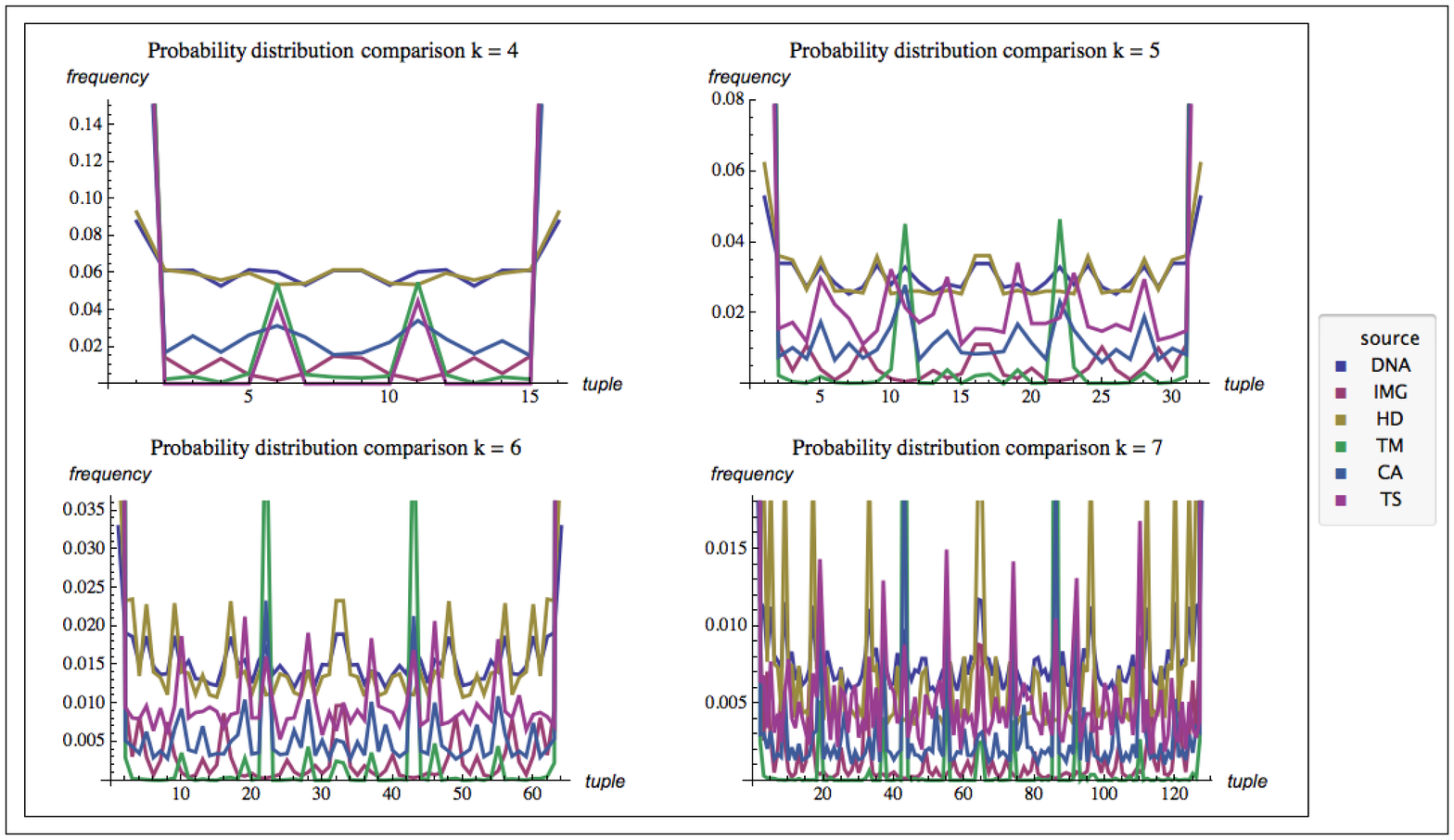}
\end{center}
\caption{Comparisons of all frequency distributions of tuples of length $k$, from physical sources and from abstract computing machines.}
\label{graph3}
\end{figure}

\subsection{The problem of overfitting}

When looking at a set of data following a distribution, one can safely claim in statistical terms that the source generating the data is of the nature that the distribution suggests. Such is the case when a set of data follows a model, where depending on certain variables, one can say with some degree of certitude that the process generating the data follows the model.

However a common problem is the problem of overfitting, that is, a false model that may fit perfectly with an observed phenomenon\footnote{For example, Ptolemy's solar system model model.}. Levin's universal distribution, however, is optimal over all distributions\cite{levin2}, in the sense that the algorithmic model is by itself the simplest possible model fitting the data if produced by some algorithmic process. This is because $m$ is precisely the result of a distribution assuming the most simple model in algorithmic complexity terms, in which the shortest programs produce the elements leading the distribution. That doesn't mean, however, that it must necessarily be the right or the only possible model explaining the nature of the data, but the model itself is ill suited to an excess of parameters argument. A statistical comparison cannot actually be used to categorically prove or disprove a difference or similarity, only to favor one hypothesis over another.

\section{Possible applications}

Common data compressors are of the entropy coding type. Two of the
most popular entropy coding schemes are the Huffman coding and the
arithmetic coding. Entropy coders encode a given set of symbols with
the minimum number of bits required to represent them. These
compression algorithms assign a unique prefix code to each unique
symbol that occurs in the input, replacing each fixed-length input
symbol by the corresponding variable-length prefix codeword. The
length of each codeword is approximately proportional to the negative
logarithm of the probability. Therefore, the most common symbols use
the shortest codes.

Another popular compression technique based on the same principle is
the run-length encoding (RLE)\footnote{Implementations in different
programming languages of the run-length encoding are available at
\emph{http://rosettacode.org/wiki/Run-length\_encoding}}, wherein
large runs of consecutive identical data values are replaced by a
simple code with the data value and length of the run. This is an
example of lossless data compression. However, none of these methods
seem to follow any prior distribution\footnote{The authors were unable
to find any reference to a general \emph{prior} image compression
distribution.}, which means all of them are a posteriori techniques
that after analyzing a particular image set their parameters to better
compress it. A sort of prior compression distributions may be found in
the so-called dictionary coders, also sometimes known as substitution
coders, which operate by searching for matches between the text to be
compressed and a set of strings contained in a static data structure.

In practice however, it is usually assumed that compressing an image
is image dependent, i.e. different from image to image. This is true
when prior knowledge of the image is available, or there is enough
time to spend in analyzing the file so that a different compression
scheme can be set up and used every time. Effectively, compressors
achieve greater rates because images have certain statistical
properties which can be exploited. But what the experiments carried
out here suggest for example is that a general optimal compressor for
images based on the frequency distribution for images can be
effectively devised and useful in cases when neither prior knowledge
nor enough time to analyze the file is available. The distributions
found, and tested to be stable could therefore be used for prior image
compression techniques. The same sort of applications for other data
sets can also be made, taking advantage of the kind of exhaustive
calculations carried out in our experiments.

The procedure also may suggest a measure of \emph{algorithmicity}
relative to a model of computation: a system is more or less
algorithmic in nature if it is more or less closer to the average
distribution of an abstract model of computation. It has also been
shown\cite{delahayezenil1} that the calculation of these distributions
constitute an effective procedure for the numerical evaluation of the
algorithmic complexity of short strings, and a mean to provide
stability to the definition--independent of additive constants--of
algorithmic complexity.
unfolding runtime.

\section{The meaning of \emph{algorithmic}}

Perhaps it may be objected that we have been careless in our use of the term \emph{algorithmic}, not saying  exactly what we mean by it. Nevertheless, \emph{algorithmic} means nothing other than what this paper has tried to convey by the stance we have taken over the course of its arguments.

In our context, \emph{Algorithmic} is the adjective given to a set of processes or rules capable of being effectively carried out by a computer in opposition to a truly (indeterministic) random process (which is uncomputable). Classical models of computation\footnote{Albeit assuming the Church-Turing thesis.} capture what an algorithm is but this paper (or what it implies) experimentally conveys the meaning of \emph{algorithmic} both in theory and in practice, attempting to align the two. On the one hand, we had the theoretical basis of algorithmic probability. On the other hand we had the empirical data. We had no way to compare one with the other because of the non-computability of Levin's distribution (which would allow us to evaluate the algorithmic probability of an event). We proceeded, however, by constructing an experimental algorithmic distribution by running abstract computing machines (hence a purely algorithmic distribution), which we then compared to the distribution of empirical data, finding several kinds of correlations with different degrees of significance. For us therefore, algorithmic means the exponential accumulation of pattern producing rules and the isolation of randomness producing rules. In other words, the accumulation of simple rules. 

Our definition of \emph{algorithmic} is actually much stronger than the one directly opposing true randomness. Because in our context something is algorithmic if it follows an algorithmic distribution (e.g. the experimental distribution we calculated). One can therefore take this to be a measure of \emph{algorithmicity}: the degree to which a data set approaches an experimentally produced algorithmic distribution (assumed to be Levin's distribution). The closer it is to an algorithmic distribution the more algorithmic.

So when we state that a process is algorithmic in nature, we mean that it is composed by simple and deterministic rules, rules producing patterns, as algorithmic probability theoretically predicts. We  think this is true of the market too, despite its particular dynamics, just as it is true of empirical data from other very different sources in the physical world that we have studied.

\section{Conclusions}
\label{conclusions}

Our findings suggest that the information in the world might be the result of processes resembling processes carried out by computing machines. That does not necessarily imply a trivial reduction more than talking about algorithmic simple rules generating the data as opposed to random or truly complicated ones. Therefore we think that these correlations are mainly due to the following reason: that general physical processes are dominated by algorithmic simple rules. For example, processes involved in the replication and transmission of the DNA have been found\cite{liz} to be concatenation, union, reverse, complement, annealing and melting, all they very simple in nature. The same kind of simple rules may be the responsible of the rest of empirical data in spite of looking complicated or random. As opposed to simple rules one may think that nature might be performing processes represented by complicated mathematical functions, such as partial differential equations or all kind of sophisticated functions and possible algorithms. This suggests that the DNA carries a strong algorithmic component indicating that it has been developed as a result of algorithmic processes over the time, layer after layer of accumulated simple rules applied over and over.

So, if the distribution of a data set approaches a distribution produced by purely algorithmic machines rather than the uniform distribution, one may be persuaded within some degree of certainty, that the source of the data is of the same (algorithmic) nature just as one would accept a normal distribution as the footprint of a generating process of some random nature. The scenario described herein is the following: a collection of different distributions produced by different data sets produced by unrelated sources share some properties captured in their frequency distributions, and a theory explaining the data (its regularities) has been presented in this paper.

There has hitherto been no way to either verify or refute the information-theoretic notion, beyond the metaphor, of whether the universe can be conceived as either the output of some computer program or as some sort of vast digital computation device as suggested by some authors\cite{fredkin, schmidhuber, wolfram, lloyd}.  

We think we've devised herein a valid statistical test independent of any bias toward either possibility. Some indications of correlations have been found having weak to strong significance. This is the case with distributions from the chosen abstract devices, as well as with data from the chosen physical sources. Each by itself turned out to show several degrees of correlation. While the correlation between the two sets was partial, each distribution was correlated with at least one distribution produced by an abstract model of computation. In other words, the physical world turned out to be statistically similar in these terms to the simulated one.

\end{document}